\documentclass[aip,pre]{revtex4-1}
\usepackage{graphicx}
\usepackage{epsfig}
\usepackage{dcolumn}
\usepackage{bm}
\usepackage{color}
\usepackage{amsthm}
\usepackage{algorithmic}
\usepackage{algorithm}

\def\(({\left(}
\def\)){\right)}
\def\[[{\left[}
\def\]]{\right]}

\def\normeq{\cong}

\newcommand{\be}{\begin{equation}}
\newcommand{\ee}{\end{equation}}
\newcommand{\bea}{\begin{eqnarray}}
\newcommand{\eea}{\end{eqnarray}}

\begin{document}

\title{
  Mean-field message-passing equations in the Hopfield model and
its generalizations}
 \author{Marc M\'ezard
}
\affiliation{
Physics department, Ecole Normale Sup\'erieure, PSL Research University, Paris. 
}

\begin{abstract}
Motivated by recent progress in using restricted Boltzmann machines as
preprocessing algorithms for deep neural network, we revisit the
mean-field equations (belief-propagation and TAP equations) in the best understood such
machine, namely the  Hopfield model of neural networks, and we
explicit how they can be used as iterative message-passing algorithms, providing
a fast method to compute the local polarizations of neurons.
 In the ``retrieval phase'' where  neurons
polarize in the direction of one memorized pattern, we point out a major
difference between the belief propagation and TAP equations : the set of
belief propagation equations depends on the pattern which is
retrieved, while one can use a unique set of TAP equations. This makes the latter method much better suited
for applications in the learning process of restricted Boltzmann
machines.
In the case where the patterns memorized in the Hopfield model are not
independent, but are correlated through a 
combinatorial structure, we show that the TAP equations have to be
modified. This modification can be seen either as an alteration of the
reaction term in TAP equations, or, more interestingly, as the
consequence of message passing on a graphical model with several
hidden layers, where the number of hidden layers depends on the depth
of the correlations in the memorized patterns. This layered structure is
actually necessary when one deals with more general restricted
Boltzmann machines.
\end{abstract}

\date{\today}

\maketitle


\newpage
\section{Introduction}

The interest into neural networks has been revived recently by a  series of
practical successes using so-called ``deep neural
networks'' to solve important and difficult problems in artificial
intelligence, ranging from image segmentation to speech recognition
(see \cite{LeCun:2015rz} and references therein). The crucial learning 
phase in these applications is often started by using techniques for unsupervised learning, like restricted
Boltzmann machines (RBM) \cite{HintonRBM2007} or auto-encoders \cite{autoencoder2008}, in order to
obtain a first set of synaptic weights that is then optimized in a
supervised learning process using back-propagation. 

The unsupervised learning in RBMs is an important 
problem. Its difficulty comes from the necessity
  to compute the correlation functions of a general spin
 systems. The correlation functions can be approximated by numerical
 methods like Monte-Carlo, but this is rather
 time-consuming. Alternative methods use local estimates of the
 correlations\cite{WellingHinton2002,Tieleman2008}, or
those that can be deduced by message-passing algorithms based on iteration of  local mean-field
equations. This last approach, which was pioneered in
\cite{KappenRodriguez1998,Tanaka1998}, has received more and more
attention recently
\cite{HuangToyoizumi2015,GabrieTramelKrzakala2015,TramelManoel_etal2016}, and 
it seems that these sophisticated message-passing algorithms can be
quite useful in RBM learning.  In recent years, message-passing has
proved successful, both for analytical studies and for algorithm design, in several important problems
of computer science including error correcting codes  (for  a reviews see for instance
\cite{RichardsonUrbanke08}), constraint
satisfaction problems (for a  review, see for instance
\cite{MezardMontanari09}), statistical inference (for a  review, see for instance
\cite{Zdeborova2015}), compressed sensing
\cite{BayatiMontanari10,Rangan10,Rangan10b,KrzakalaPRX2012,KrzakalaMezard12},
or learning in perceptrons \cite{Mezard1989,Baldassi2007,Baldassi2015}.

The aim of this paper is to revisit the mean-field equations, and
their use as message-passing algorithm, in the Hopfield model of
neural networks\cite{Hopfield82}. The Hopfield model, a model of binary neurons interacting by pairs, with
synaptic weights chosen in such a way that the neurons tend to
polarize spontaneously towards one of the memorized ``patterns'', can
also be seen as a RBM. It is in fact one of the best understood models of neural
networks and of RBMs, and it provides
an excellent starting point to understand the mean-field message-passing equations and their possible use as algorithms.

The present paper addresses four issues. The first one is the derivation
of the various types of mean-field equations in the Hopfield model,
the second one is their use as an algorithm, the third one is an
analysis of the mean-field equations in a generalized Hopfield model
where patterns have a combinatoric type of correlation. The fourth one
is the
generalization of the whole approach to RBMs which are of a more
general type than the Hopfield model. 
 
It is useful
to clarify the mean-field equations in the case of the Hopfield model because
several forms of these equations exist, under various names and
acronyms:
\begin{itemize}
\item
Belief
Propagation: BP
\item 
Relaxed Belief Propagation: rBP
\item
Thouless-Anderson Palmer equations: TAP
\item 
Approximate Message Passing: AMP
\item
Generalized Approximate Message Passing: GAMP
\end{itemize}

We shall see that each version is useful: BP and rBP form the basis of
the statistical analysis called the cavity method \cite{MPV87} (also
known as state evolution or density evolution in the recent computer-science litterature) which gives
the phase diagram of this problem. They can be
used to derive TAP equations \cite{TAP}, which are also called AMP
equations in the recent computer science litterature.
TAP equations were originally
derived in the Hopfield model by \cite{MPV87}. Through our derivation
of TAP equations as simplifications of the general BP equations
(related to the one done in
\cite{KabashimaSaad2001}), we confirm the validity of these equations,
in spite of previous claims by 
\cite{Nakanishi1997} and \cite{ShamirSompolinsky2000} that they were
incorrect. All methods actually lead to the same TAP equations as \cite{MPV87}.

An important point which is clarified in the present approach concerns
the use of message-passing mean-field equations in the ``retrieval
phase'' of the Hopfield model, the phase where the neurons polarize
spontaneously in the direction of one of the stored patterns (and where
the model can be used as an associative memory). In this phase, the
usual simplification of BP equations into rBP,  which  assumes that messages
have a Gaussian distribution, is incorrect and one must treat separately some of the messages which are
associated with the specific pattern where the polarization
develops. The equivalent of the rBP equations, taking into account
this modification (called rBP-M in the following), are structurally
distinct from the usual rBP equations.  However this distinction
disappears when one writes TAP equation. This makes the
TAP equations much better suited for algorithmic applications.

The Hopfield model is a system of binary neurons (or Ising spins),
with infinite range pairwise interactions. It is thus intimately
related to the infinite range model of spin glasses of Sherrington and
Kirkpatrick (SK) \cite{SK}, but it differs from it in the detailed
structure of the interactions between spins. Instead of being
independent random variables, the coupling constants between the spins
are built from a set of predetermined patterns that one wants to
memorize. This structure leads to a modification of the Onsager
reaction term in the TAP equations. Our derivation shows that this
modification is easily understood by using a representation of the
Hopfield model with two layers, a layer of visible neuron-variables,
and a layer of hidden pattern-variables. The exchange of messages
between these two layers (in which the Hopfield model is seen as a
RBM) precisely leads to the modification of the Onsager reaction
term. We will show that this structure can actually be iterated. We
define a modified Hopfield model where the patterns are not
independent random variables, but they are built by combinations of
more elementary objects, called features. In this case, we show that the TAP equations
can be understood by a neural network  with three layers, in which one
adds, between the layer of visible neuron-variables and the layer of
hidden pattern-variables, another layer of hidden
feature-variables. This spontaneous emergence of more hidden layers
when one handles a more structured type of problem is interesting in
itself: one might hope that it could lead to an explanation of the
success of multilayered network and deep learning in practical tasks
where the information certainly contains  a deep hierarchy of
combinatorial correlations. 

We do not address here the full problem of learning in the Hopfield model
or in RBMs. We only study the ``direct'' problem of determining the
polarization of each neuron (from which one can deduce the pair
correlations by using linear response). However a good control of this direct
problem is an essential ingredient of most unsupervised learning protocols.

The paper is organized as follows:

- Section 2 provides basic definitions of the Hopfield model and
recalls its phase diagram.

- Section 3 derives the mean field equations. It starts with the
phases where there is no spontaneous polarization of the neurons, and
derives successively  the BP
equations, their rBP simplification using Gaussian messages, and
finally the TAP (or AMP) equations. It then studies the modifications of these equations when one
works in the retrieval phase. The consistency of the BP equations with the
standard replica results (a consistency which had been disputed in
\cite{ShamirSompolinsky2000}) is then explicitly shown.

- Section 4 explains how the mean-field equations can be turned into
algorithms, by iterating them with a careful update schedule.

- Section 5 studies a modified Hopfield model in which the patterns
are no-longer independent, but they are built as combinations of more
elementary random variables. We work out the modification of BP and
TAP equations in this case, using a representation of the problem with
two layers of hidden variables, on top of the layer of visible neurons.

- Section 6 derives the message-passing algorithms obtained from mean-field equations (BP, rBP and TAP) in a
general model of RBM. 

- Section 7 provides some concluding remarks, and perspectives for
further studies.

\section{The Hopfield model}
\label{sec:Hopfield}
\subsection{Definitions}
In the Hopfield model \cite{Hopfield82}, neurons are modeled as $N$ binary spins $s_i$,
$i=1,\dots,N$, taking values in $\{\pm 1\}$.
These spins interact by pairs, the energy of a spin configuration is
\be
E=-\frac{1}{2}\sum_{i,j} J_{ij} s_i s_j\ .
\label{ene_Hopf}
\ee
This is a spin glass model where the coupling constants $J_{ij}$ take
a special form. Starting from $p$ ``patterns'',
which are spin configurations 
\be
\xi_i^\mu =\pm 1,\ i\in\{1,\dots n\},\ \mu\in\{1,\dots p\}\ ,
\ee
the coupling constants are defined as
\be
J_{ij}=\frac{1}{N}\sum_\mu \xi_i^\mu \xi_j^\mu\ .
\label{Jdef}
\ee
Given an instance defined by the set of couplings $J=\{J_{ij}\}$, the
Boltzmann distribution of the spins, at inverse temperature $\beta$,
is defined as
\be
P_J(s)= \frac{1}{Z} e^{(\beta/2) \sum_{i,j} J_{ij} s_i s_j} \ .
\ee
Using a Gaussian transformation, the partition function $Z$ can be re-written as
\be
Z= \sum_s \int \prod_\mu \frac{d \lambda_\mu}{\sqrt{2\pi\beta}}\ 
\exp\left[
-\frac{\beta}{2}\sum_\mu\lambda_\mu^2+\beta
  \sum_{\mu,i}\frac{\xi_i^\mu}{\sqrt{N}} s_i \lambda_\mu
\right]\ .
\label{Z_def}
\ee
This expression shows that the Hopfield model is also a model of $N$
binary spins $s_i$ and $P$ continuous variables with a Gaussian measure,
$\lambda_\mu$, interacting through random couplings
$\xi_i^\mu / \sqrt{N}$ which are iid random variables taking
values $\pm 1/\sqrt{N}$ with probability $1/2$. This is nothing but a Restricted
Boltzmann Machine in which the visible neurons are binary variables
that  interact with $P$ 
hidden continuous variables with a Gaussian distribution. The
variable $\lambda_\mu$ can be interpreted as the projection of the spin
configuration on the pattern $\mu$, as suggested by the identity
relating its mean $\langle \lambda_\mu \rangle$ and the expectations
values of the spins:
\be
\langle \lambda_\mu \rangle = \frac{1}{\sqrt{N}} \sum_i \xi_i^\mu
\langle s_i \rangle\ .
\ee

\subsection{Known results}
The phase diagram of the Hopfield model has been studied in detail in
\cite{AGS85-1}\cite{AGS85-2} and subsequent papers. In the thermodynamic limit where the number of
neurons $N$ and the number of patterns $P$ go to infinity with a fixed
ratio $\alpha=P/N$, the phase diagram is controlled by the temperature
$T=1/\beta$ and the ratio $\alpha$. One finds three main phases:
\begin{itemize}
\item The paramagnetic phase. At high enough temperatures,
  $T>T_g(\alpha)$, the spontaneous polarization of each neuron vanishes
  $\langle s_i \rangle=0$.
\item The retrieval phase. In a regime of low enough temperature and
  low enough $\alpha$, there exists a retrieval phase, where the
  neurons have a spontaneous polarization in the direction of one of
  the stored patterns $\mu$. This means that, in the thermodynamic
  limit :
\bea
\frac{1}{N} \sum_i \langle s_i \rangle \xi_i^{\mu}&=&M
\\
\frac{1}{N} \sum_i \langle s_i \rangle \xi_i^{\nu}&=&0 \ \
( \nu \neq \mu )
\eea
For symmetry reasons, there exist two retrieval states one with a
polarization $M>0$ (where $M$ is a function of $\alpha,\beta$), one
with the polarization $-M$ (pointing opposite to the pattern).

The transition corresponding to the appearance of retrieval states is
a first order transition. One should thus distinguish two
temperatures: at  $T<T_M(\alpha)$, retrieval states first appear as
metastable states, at a lower temperature $T<T_c(\alpha)$, they become global minima of
the free energy.
\item
The spin glass phase. In an intermediate range of temperature, or at
large $\alpha$, the neurons acquire a spontaneous polarization, but
in some directions which are not in the direction of one of the
patterns. In the spin glass phase, the spin-glass order parameter $q$,
defined by
\be
q=\frac{1}{N} \sum_i \langle s_i \rangle^2
\ee
is strictly positive, while
\be
\forall \mu \ :\ \ \frac{1}{N} \sum_i \langle s_i \rangle \xi_i^{\mu} =0
\ee
\end{itemize}

\begin{figure*}[!ht]
\begin{center}
\hspace{-0.35cm}
\includegraphics[width=0.95\textwidth]{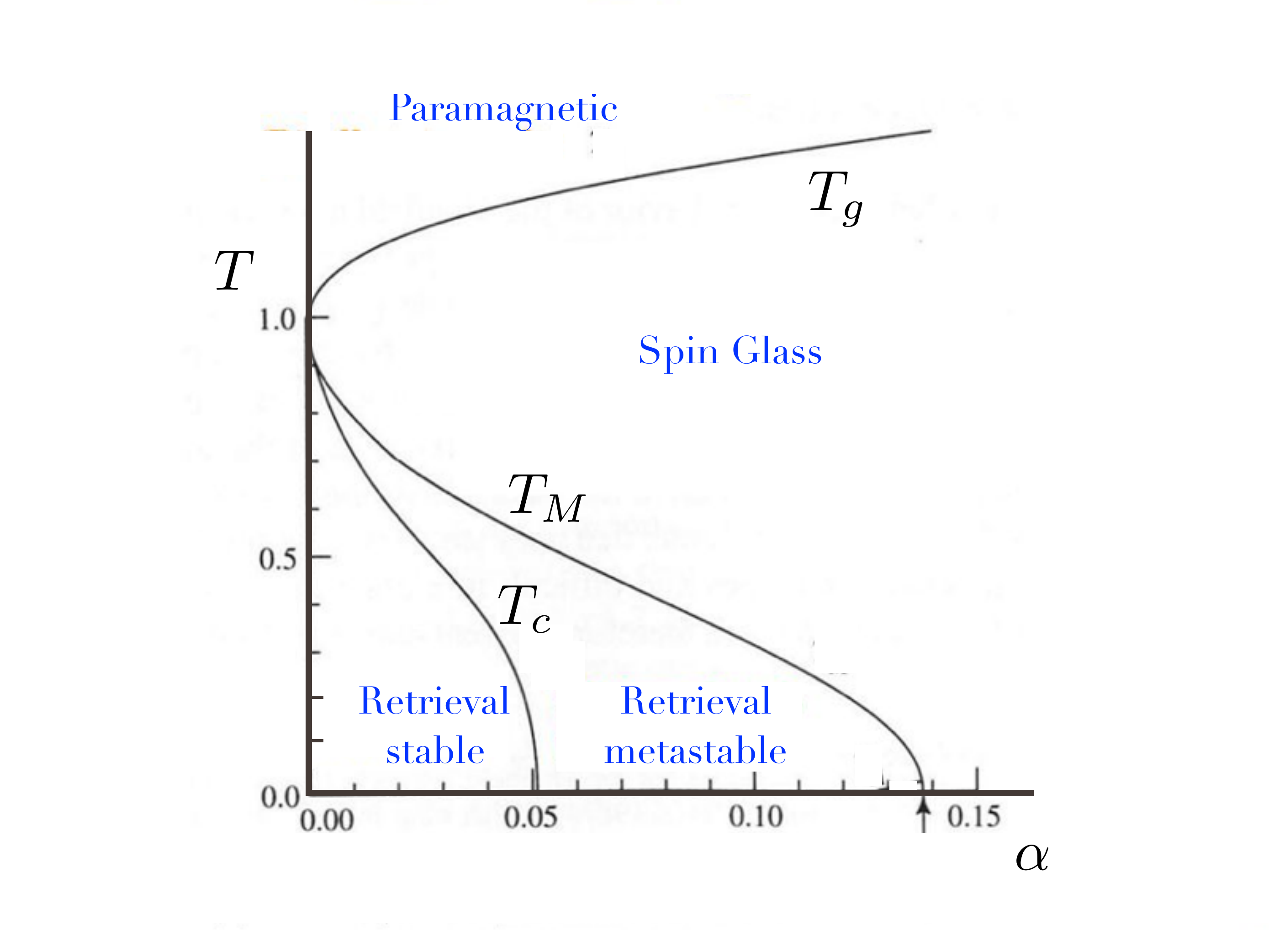}
\caption{Phase diagram of the Hopfield model, from \cite{AGS85-2}}
\label{fig:phase-diag}
\end{center}
\end{figure*}

The phase diagram is recalled in Fig. \ref{fig:phase-diag}. 
\section{Mean-field equations}

\subsection {Belief propagation}
\begin{figure*}[!ht]
\begin{center}
\hspace{-0.35cm}
\includegraphics[width=0.95\textwidth]{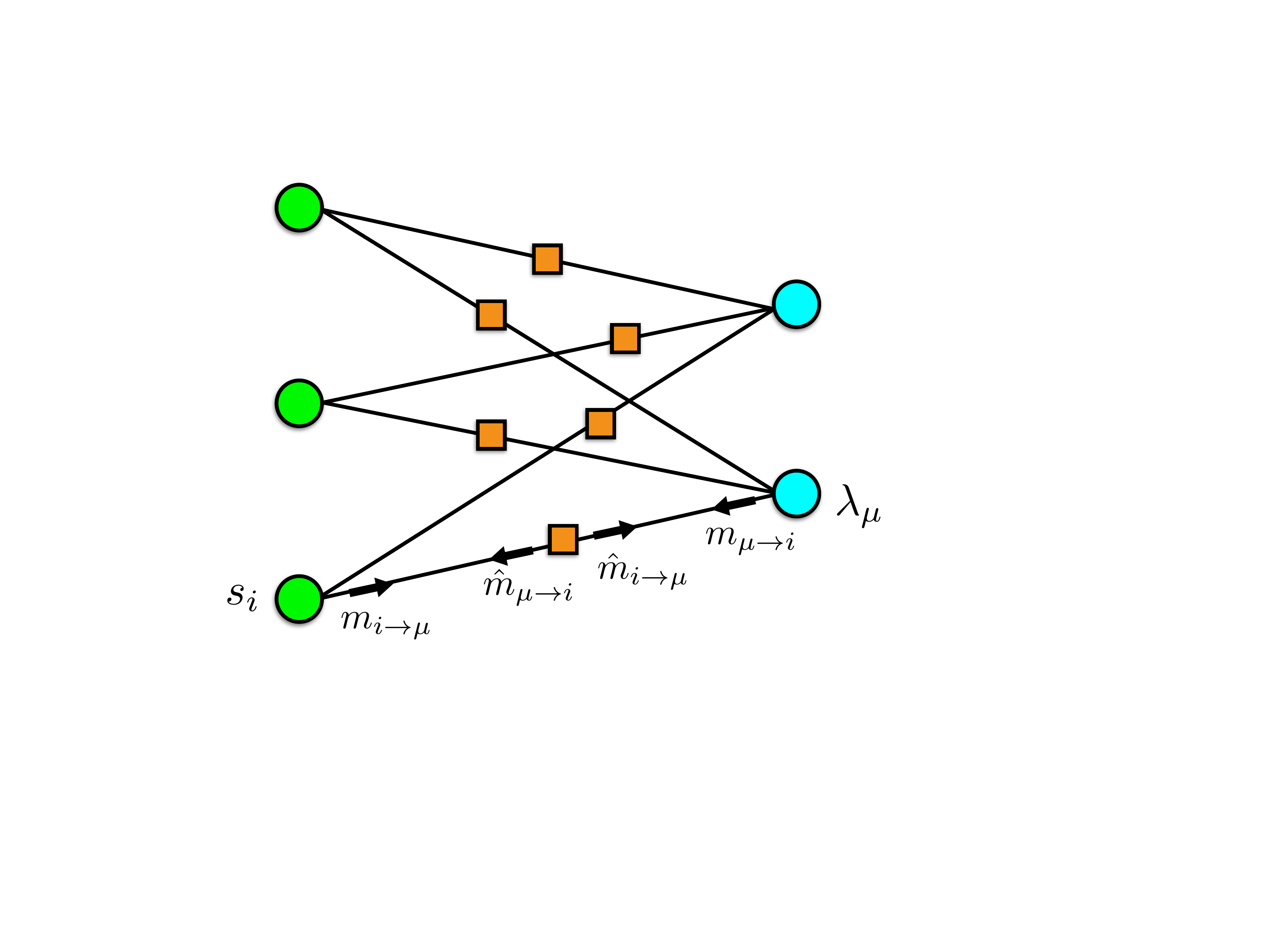}
\caption{Factor graph of the Hopfield model, in the representation
  using spin-variables ($s_i$, left circles) and
  pattern-variables ($\lambda_\mu$, right circles). For each pair of  spin-variable
   and pattern-variable, there is an interaction factor (squares). The
   graph also defines the messages used in belief propagation}
\label{fig:factorgraph}
\end{center}
\end{figure*}

We use the representation (\ref{Z_def}) of the Hopfield
model. Figure \ref{fig:factorgraph} shows the factor graph for this
problem. The BP equations are written using the
standard procedure (see for instance \cite{MezardMontanari09}).
For the
Hopfield model, this approach was first used by
\cite{KabashimaSaad2001}. The BP equations are expressed
 in terms of
four types of messages: $m_{i\to\mu}(s_i)$, $m_{\mu\to i}(\lambda_\mu)$,
$\hat m_{\mu\to i}(s_i)$, $\hat m_{i\to\mu}(\lambda_\mu)$. The
message $m_{i\to\mu}(s_i)$, being a probability of a binary variable,
can be parameterized in terms of a single number $h_{i\to\mu}$,
denoted 
``cavity field'',
defined by:
\be
m_{i\to\mu}(s_i) \normeq e^{ \beta h_{i\to\mu} s_i} \ ,
\ee
(in this paper, the symbol `$\normeq$' denotes equality up to a
constant :
if $p(\,\cdot\,)$ and $q(\,\cdot\,)$ are two measures 
on the same space -not necessarily normalized-,
we write $p(x)\normeq q(x)$ if there exists $C>0$ such that $ p(x) =
C\, q(x)$).
Similarly:
\be
\hat m_{\mu\to i}(s_i) \normeq e^{ \beta \hat h_{\mu\to i} s_i }\ .
\ee

The BP equations are:
\bea
h_{i\to\mu} &=& \sum_{\nu(\neq \mu)} \hat h_{\nu \to i}  \label{BP1}\\
\hat m_{\mu \to i}  (s_i) &\normeq&  \int d\lambda_\mu 
m_{\mu\to i}(\lambda_\mu) \exp\left(  (\beta/\sqrt{N}) \xi_i^\mu s_i
  \lambda_\mu\right)\label{BP2}\\
\hat m_{i\to\mu}(\lambda_\mu)&\normeq& \cosh\beta\left(
  h_{i\to\mu}+(\xi_i^\mu/\sqrt{N})\lambda_\mu\right)\label{BP3}\\
m_{\mu\to i}(\lambda_\mu) &\normeq&e^{-(\beta/2) \lambda_\mu^2}\prod_{j(\neq
  i)} \hat m_{j\to \mu}(\lambda_\mu)\label{BP4}
\eea
Notice that in this notation the message $\hat m_{i\to \mu}$ is not a
normalizable distribution of $\lambda_\mu$. This is not a problem
because it appears in the equations only through the combination
(\ref{BP4}) which is normalizable thanks to the presence of the local
Gaussian measure $e^{-(\beta/2) \lambda_\mu^2}$.

\subsection{Relaxed BP equations}
The general BP equations are not very useful because the messages
$m_{\mu \to i}$ and $\hat m_{i \to \mu}$ are functions of the
continuous variables $\lambda_\mu$. However the messages can be simplified,
by noticing that, in the thermodynamic limit,  they are
actually Gaussian distributions and can be parameterized by just two
moments. This simplification leads to equations that are 
usually called ``relaxed BP'' (rBP) in the litterature. 
It was first used in the cavity method for systems with long-range interactions
(\cite{MPV87}, and has been developped in various problems of
communication theory
\cite{Kabashima2003,Tanaka2005,Montanari2006,GuoWang06}. 

We shall first work out this simplification in the phase where there is
no condensation on any pattern. Technically, this means that the
distributions  $m_{\mu\to i}(\lambda_\mu)$ are dominated by values of
$\lambda_\mu$ which are finite (in the large $N$ limit). It is easy to
see that, in this case, (\ref{BP3}) can be expanded around
$\lambda_\mu/\sqrt{N}=0$ and  the BP equations (\ref{BP1}-\ref{BP4}) close
under the hypothesis that the messages $m_{\mu\to i}(\lambda_\mu)$ are
Gaussian distributions.

Under
this assumption, one can parameterize these messages in terms of
their two first moments. This leads to the so-called rBP equations. 
We define:
\bea
a_{\mu\to i} &= & \int d\lambda_\mu 
m_{\mu\to i}(\lambda_\mu) \lambda_\mu \label{adef}\\
c_{\mu\to i} &= & \int d\lambda_\mu 
m_{\mu\to i}(\lambda_\mu) \lambda_\mu^2- a_{\mu\to i}^2\label{cdef}
\eea
In order to derive the rBP equations, we first derive the asymptotic
form of  the function $ \hat m_{i\to\mu} (\lambda_\mu)$ in the large
$N$ limit:
\bea
\hat m_{i\to\mu} (\lambda_\mu) = \exp\left( 
\beta\frac{\xi_i^\mu }{\sqrt N}
  \lambda_\mu \tanh(\beta h_{i\to\mu})+
\frac{\beta^2}{2N}
  \lambda_\mu^2[1-\tanh^2(\beta h_{i\to\mu})]
\right)
\eea
Inserting this expression into (\ref{BP4}) we get:
\bea
c_{\mu \to i} &=& \frac{1}{\beta}\frac{
1
}
{
1-(\beta/N)\sum_{j(\neq i)}  [1-\tanh^2(\beta h_{j\to\mu})]
}\\
a_{\mu \to i} &=& \frac{1}{\sqrt N} \frac{
 \sum_{j(\neq i)} \xi_{j}^\mu \tanh(\beta h_{j \to \mu})
}
{
1-(\beta/N)\sum_{j(\neq i)}  [1-\tanh^2(\beta h_{j\to\mu})]
} \label{rBP_a}
\eea
Eqs. (\ref{BP1}) and (\ref{BP2}) can be re-written as:
\bea
h_{i\to \mu} = \sum_{\nu(\neq \mu)} \frac{\xi_i^\nu}{\sqrt{N}} a_{\nu
  \to i}
\label{rBP_h}
\eea
Equations (\ref{rBP_a}) and (\ref{rBP_h}) form a set of $2 N P$
equations for the $2NP$ variables $a_{\mu \to i}$ and $h_{i \to \mu }$. These are the rBP equations for the Hopfield model.

\subsection{TAP (or AMP) equations}
The rBP equations relate messages propagated along
the edges of the factor graph (in the language of spin glasses they
are ``cavity-equations''). In the large $N$ limit it is possible, and very useful for algorithmic
purpose, to simplify these rBP equations and turn them into a set of
equations which relate ``site''-quantities associated with the variable-nodes
of the factor graph. This allows to go from $2NP$ variables to $N+P$
variables, and leads to an effective reduction of computer time and
memory. The equations that relate them are analogous to those found in
\cite{TAP} for spin glasses, hence the name TAP equations. In
computer science, they are often called approximate-message-passing
(AMP) equations \cite{DonohoMaleki09,
BayatiMontanari10,Rangan10b,SCHNITER,KrzakalaPRX2012,SCHNITER-BIG,KrzakalaMezard12}. To
avoid confusion, notice
that, in
their paper on the Hopfield model  \cite{KabashimaSaad2001}, Kabashima
and Saad use the same word (TAP equations) both for the rBP equations
and what we call TAP equations. We prefer to use two different terms,
in line with the terminology which is presently most common in
information theory.

The site-variables are local fields defined as:
\bea
H_{i} &=& \sum_{\nu} \frac{\xi_i^\nu}{\sqrt{N}} a_{\nu
  \to i}
\label{Hdef}
\\
A_{\mu} &=&
\frac{1}{\sqrt N} \frac{
 \sum_{j} \xi_{j}^\mu \tanh(\beta h_{j \to \mu})
}
{
1-(\beta/N)\sum_{j}  [1-\tanh^2(\beta h_{j\to\mu})]
} 
\label{Adef}
\eea
They give the expectation values of the variables:
\bea
\langle s_i\rangle&=&M_i= \tanh(\beta H_i)\\
\langle \lambda_\mu\rangle&=&A_\mu
\eea
We shall derive here a closed set of $N+P$ equations that relate these
$N+P$ variables. 

The main idea of the derivation of TAP equations comes from the
observation that the rBP message $a_{\mu \to i}$ should be nearly
equal to $A_{\mu}$, up to small corrections that can be handled
perturbatively in the large $N$ limit. Similarly, $h_{i \to \mu}$ is
nearly equal to $H_i$, up to small corrections. Let us work out the
explicit form of these corrections. We define 
\be
q=\frac{1}{N}\sum_i \tanh^2 (\beta H_i)
\label{qdef}
\ee
We first notice that
\be
h_{j \to \mu} =H_j-\frac{\xi_j^\mu}{\sqrt N}a_{\mu \to j}
\ee
Therefore:
\be 
\frac{1}{N} \sum_{j}  [1-\tanh^2(\beta h_{j\to\mu})] \simeq 1-q 
\ee
up to corrections which vanish when $N\to\infty$. Therefore :
\bea
A_\mu \simeq \frac{1}{1-\beta(1-q)} \frac{1}{\sqrt N} 
 \sum_{j} \xi_{j}^\mu \tanh(\beta h_{j \to \mu})
\eea
and:
\bea
a_{\mu \to i}\simeq A_\mu-\frac{1}{1-\beta(1-q)} \frac{1}{\sqrt N} 
 \xi_{i}^\mu \tanh(\beta h_{i \to \mu})
\eea
In this last expression, the second term is a correction of order
$1/\sqrt{N}$. In this correction we can substitute $h_{i \to \mu}$ by
$H_i$, the difference would give a contribution of order
$O(1/N)$ to $a_{\mu \to i}$, which can be neglected.
Therefore :
\bea
a_{\mu \to i}\simeq A_\mu-\frac{1}{1-\beta(1-q)} \frac{1}{\sqrt N} 
 \xi_{i}^\mu \tanh(\beta H_i) 
\eea
Substituting this expression in the definition (\ref{Hdef})  of $H_i$
we get:
\bea
H_i\simeq \sum_{\nu} \frac{\xi_i^\nu}{\sqrt{N}}
A_\nu-\frac{\alpha}{1-\beta(1-q)} \tanh(\beta H_i)
\label{TAP1}
\eea

Considering now the definition (\ref{Adef}) of $A_\mu$ we can expand
it as 
\bea
A_\mu &=& \frac{1}{1-\beta(1-q)} 
 \sum_{j} \frac{\xi_{j}^\mu}{\sqrt N}  \tanh\left [\beta \left(H_j -\frac{\xi_j^\mu}{\sqrt N}
 a_{\mu \to j}\right)\right]\nonumber \\
&\simeq&
\frac{1}{1-\beta(1-q)} 
 \sum_{j} \frac{\xi_{j}^\mu}{\sqrt N}\tanh\left[\beta \left(H_j -\frac{\xi_j^\mu}{\sqrt N}
 A_\mu\right) \right] +O(1/N)\nonumber\\
&\simeq&
\frac{1}{1-\beta(1-q)} 
 \sum_{j} \frac{\xi_{j}^\mu}{\sqrt N}\left[\tanh(\beta H_j )-\beta
   \frac{\xi_j^\mu}{\sqrt N}(1-\tanh^2(\beta H_j) A_\mu\right]
\eea
This gives:
\bea
A_\mu=\frac{1}{\sqrt N} \sum_{j} \xi_{j}^\mu \tanh(\beta H_j )
\label{TAP2}
\eea

Equations (\ref{TAP1}) and (\ref{TAP2}), together with the definition
(\ref{qdef}), are the TAP (or AMP) equations which relate the $N+P$
variables $H_i$ and $A_\mu$. 
It turns out that they are linear in $A_\mu$, and these variables can
thus be eliminated (notice however that this is a specific feature of
the Hopfield model, due to the Gaussian nature of variables
$\lambda_\mu$: as we shall see in Sect. 6, this is no longer true for
more general RBMs, where the measure on $\lambda_\mu$ is non-Gaussian).
Eliminating $A_\mu$, we write closed
equations for the $N$ local fields $H_i$
\bea
H_i=\frac{1}{N} \sum_j J_{ij} \tanh(\beta H_j) -
\frac{\alpha}{1-\beta(1-q)} \tanh(\beta H_i)
\eea
An alternative presentation of these TAP equations are in terms of the
local magnetizations $M_i=\tanh(\beta H_i)$:
\bea
M_i &=&\tanh\left[ \beta \sum_j J_{ij} M_j -
\frac{\alpha \beta}{1-\beta(1-q)} M_i \right] \label{TAPmag1}\\
&=&\tanh\left[ \beta \sum_{j(\neq   i)}  J_{ij} M_j -
\frac{\alpha \beta^2 (1-q)}{1-\beta(1-q)} M_i \right] \label{TAPmagNT}\\
q&=&\frac{1}{N}\sum_i M_i^2 \label{TAPmag2}
\eea
These TAP equations were first derived in \cite{MPV87} using the cavity method. The re-derivation that we have presented
here uses a different approach, namely the BP equations and their simplification at large
$N$, and obtains the same result. 

The claims in \cite{Nakanishi1997} and \cite{ShamirSompolinsky2000}
according to which these equations are wrong were probably based on
their misunderstanding of the presence of diagonal terms in
(\ref{TAPmag1}). Actually the TAP equations that they derive agree
with ours, and with the original finding in \cite{MPV87}, as can be seen explicitly in the form (\ref{TAPmagNT}).

While \cite{MPV87} claimed (without writing the proof) that the TAP equations  (\ref{TAPmag2})
reproduce the known equilibrium properties of the Hopfield model found
with replicas, it was stated in  \cite{Nakanishi1997} and \cite{ShamirSompolinsky2000}  that they do
not give the well known value of the spin glass transition
temperature $T_c$, and that they disagree with the result of the replica
method. These statements are not correct. We
provide below the explicit proof that our rBP and TAP equations are in perfect
agreement with the replica result, and therefore with the known value
of $T_c$, as stated in \cite{MPV87}. The following derivation also gives a
useful pedagogical example of how the equilibrium results can be obtained
from the mean-field equations : the critical temperature can be
analysed through a study of the TAP equations, while the
replica result for the order parameter can be obtained from a
statistical analysis of the rBP equations. 

\subsection{rBP and TAP (AMP) equations in the retrieval phase}
Let us work out the modifications that take place in the retrieval
phase, when the measure
condenses on one pattern (a similar analysis can be carried out easily
in the mixed phase where the condensation takes place on a finite
number of patterns, we shall keep here to the retrieval phase). In the
retrieval phase corresponding to pattern $\mu=1$, one expects that the
distribution of $\lambda_1$ will be dominated by values close to
$\lambda_1=M \sqrt{N}$.
When deriving BP equations, the message $\hat m _{i \to
  1}(\lambda_1)$ takes the form:
\be
\hat m_{i\to1}(M)= \cosh\beta\left(
  h_{i\to 1}+M \ \xi_i^1\right)\ .
\ee
Therefore :
\bea
m_{1 \to i}(M) \normeq
e^{N \psi_{1 \to i}(M)}\ ,
\eea
where
\be
\psi_{1 \to i}(M)=-\frac{\beta}{2} M^2 + \frac{1}{N}\sum_{j(\neq i)} \log\cosh\left[\beta
  \left(h_{j\to 1}+ M \ \xi_j^1\right)\right]
\ee

In the large $N$ limit, the measure $m_{1 \to i}(M)$  is dominated by the maximum of
the function $\psi_{1 \to i}(M)$. One should notice that in the large
$N$ limit this function converges to 
\be
\psi(M)=-\frac{\beta}{2} M^2 + \frac{1}{N}\sum_{j} \log\cosh\left[\beta
  \left(h_{j\to 1}+ M \ \xi_j^1\right)\right]
\ee
The maximum of $\psi(M)$ can be either in $M=0$ or in
$M=\pm M^*$, where $M^*$ is the largest solution of the equation 
\be
M= \frac{1}{N}\sum_{j} \xi_j^1 \tanh\left[\beta
  \left(h_{j\to 1}+ M\xi_j^1 \right)\right]
\label{M_eq}
\ee
The retrieval phase is the phase where the maximum is obtained at
$M=\pm M^*$. In this case the rBP equations (\ref{rBP_a}-\ref{rBP_h})
are modified, because the messages $m_{1 \to i}$, instead of being
Gaussian distributions with finite means and variances, become
dominated by values of $\lambda_1$ close to $M\sqrt{N}$. The new set
of equations obtained in this regime will be denoted rBP-M 
(for relaxed belief propagation - magnetized) equations.

\bea
a_{\mu \to i} &=& \frac{1}{\sqrt N} \frac{
 \sum_{j(\neq i)} \xi_{j}^\mu \tanh(\beta h_{j \to \mu})
}
{
1-(\beta/N)\sum_{j(\neq i)}  [1-\tanh^2(\beta h_{j\to\mu})]
} \ \ \ , \ \ \ \mu\ge 2
\label{rBP_ret_a}
\\
h_{i\to \mu} &=& \sum_{\nu(\neq \mu,1)} \frac{\xi_i^\nu}{\sqrt{N}} a_{\nu
  \to i}+M\ \xi_i^1 \ \ \ , \ \ \ \mu\ge 2
\label{rBP_ret_hmu}
\\
h_{i\to 1} &=& \sum_{\nu(\neq 1)} \frac{\xi_i^\nu}{\sqrt{N}} a_{\nu
  \to i}
\label{rBP_ret_h1}
\eea
The rBP-M equations in the retrieval phase with condensation on pattern
$1$ are given by (\ref{M_eq}-\ref{rBP_ret_h1}).

 It should be noticed
that they involve a completely different estimate for the message
$a_{1 \to i}$ when compared to the rBP equations without
condensation. In particular, they cannot be obtained from
(\ref{rBP_a},\ref{rBP_h}) by just assuming that $a_{1\to i}$ becomes
of order $\sqrt{N}$ (such a procedure is unable to reproduce eqn(\ref{M_eq})) . 
The reason is that the condensation is a first
order transition, and the rBP equations in the retrieval phase
correspond to a solution $M>0$ to Eq.(\ref{M_eq}) that is different
from the usual one with $M=0$ (in which case one needs to consider the
$O(1/\sqrt{N})$ corrections as in (\ref{rBP_a})). The main drawback of
these rBP-M equations is that one must use a different set of
equations depending on the pattern towards which the system polarizes. This
is quite inefficient for algorithmic applications : if one does not
know a priori which pattern is being retrieved, one should run in
parallel $P=\alpha N$ different algorithms, each one testing the
possible polarization towards one of the patterns, and compare the results.

Fortunately, the situation is much better when considering TAP equations.
It is straightforward to go from these rBP-M equations to the
TAP (or AMP) equations for the retrieval phase.
One gets:

\bea
H_i &=&\sum_{\nu\ge 2} \frac{\xi_i^\nu}{\sqrt{N}}
A_\nu-\frac{\alpha}{1-\beta(1-q)} \tanh(\beta H_i)+ M\ \xi_i^1
\label{TAP_ret_1}
\\
A_\mu&=&\frac{1}{\sqrt N} \sum_{j} \xi_{j}^\mu \tanh(\beta H_j )\ \ \ ,
\ \ \mu \ge 2
\label{TAP_ret_2}
\\
M&=&\frac{1}{N}\sum_j \xi_j^1 \tanh(\beta H_j)
\eea
It turns out that these TAP equations are exactly the ones
that would be obtained from the usual
TAP equations (\ref{TAP1},\ref{TAP2}), assuming that $A_1=\sqrt{N} M$. This is
rather remarkable considering the fact that the rBP-M equations in the
retrieval phase cannot
be obtained continuously from the rBP equations without retrieval (because of the first
order phase transition discussed above). The discontinuity in the set
of rBP  equations when going from the uncondensed to the retrieval phase
thus disappears when one uses instead the TAP (GAMP) equations. This
makes the TAP equations a much better choice for algorithmic
applications.

\subsection{Consistency with the replica results}
\subsubsection{Critical temperature}
The paramagnetic solution of the TAP equations
(\ref{TAPmag1},\ref{TAPmag2}) is the solution with zero local
magnetizations, $\forall i \ : \ M_i=0$. The spin glass
transition is a second order phase transition, therefore its temperature
$T_c=1/\beta_c$ is the largest temperature where a solution with
non-zero local magnetization exists. It can be found by linearizing
the TAP equations (\ref{TAPmag1},\ref{TAPmag2}) and identifying their
instability point. Explicitly, the linearization gives:
\bea
M_i = \beta \sum_j J_{ij} M_j -
\frac{\alpha \beta}{1-\beta} M_i +O(M^3)
\eea
The direction of instability is the one of the eigenvector of the $J$
matrix with
largest eigenvalue. Denoting by $\lambda_{\tt max}$ this largest
eigenvalue, the value of $\beta_c$ is given by:
\be
1=\beta_c \lambda_{\tt max} -\frac{\alpha \beta_c}{1-\beta_c}
\label{betac_eq}
\ee

By definition, $\lambda_{\tt max}$ is the largest eigenvalue of the
matrix $N\times N$ matrix $J=(1/N) \xi \xi^T$, where the $N\times P$
matrix $\xi$ has iid random entries taking values $\pm 1$ with
probability $1/2$. In fact  the distribution of the largest eigenvalue of $J$
concentrates around
\be
\lambda_{\tt max}= (1+\sqrt\alpha)^2 \ .
\label{lmax}
\ee
This result can  be  derived using the replica method or the cavity
method. An
easy way to obtain it is to realize that the value of $\lambda_{\tt
  max}$ depends only on the first two moments of the distribution of
the matrix elements $\xi_i^\mu$. In particular it is the
same as the one which would be obtained if the entries of $\xi$ were iid with a normal
distribution of mean $0$ and variance $1$. This last case is very well
known since the work of Marcenko and Pastur \cite{MarPas}, and it
gives the value of $\lambda_{\tt max}$ written in (\ref{lmax}). 

Using (\ref{lmax}), the value of $\beta_c$ obtained from (\ref{betac_eq})
is
\be
\beta_c= \frac{1}{1+\sqrt{\alpha}}\ .
\label{betac}
\ee
This agrees with the well known result of \cite{AGS85-2} for the critical
temperature: $T_c = 1+\sqrt{\alpha}$.

\subsubsection{Order parameter}
\label{OP}
The cavity or BP equations can be used in two
distinct ways: on a single instance they can be solved by iteration,
and if a fixed point is found, this idea may be used as  an algorithm for
estimating the local magnetizations. But in the case where the
instances are generated from an ensemble (like the case that we study
here, where the $\xi_i^\mu$ are iid random variables), one can also perform
a statistical analysis of the equation.  This is the essence of the
cavity method, and is also known in the litterature on message passing
as the density evolution.

We will show that this statistical analysis of the cavity
equations give the same results as the replica method, as claimed in 
\cite{MPV87}, and contrary to the statements of
\cite{ShamirSompolinsky2000}. For simplicity, we keep here to the ``replica
symmetric'' approximation.

We start from the rBP equations. Considering first the equation
(\ref{rBP_h}) giving the cavity field $h_{i\to\mu}$, we notice that,
as the variables $\xi_i^\mu $ are iid, provided that the correlations
of the messages $a_ {\nu \to i}$ are small enough (this is the essence
of the replica symmetric approximation - see \cite{MPV87} and
\cite{MezardMontanari09}), the cavity  field $h_{i\to\mu}$ has a gaussian
distribution with mean $0$ and a variance which is independent of the
indices $i$ and $\mu$ and that we denote by
$\overline{h^2}$. Similarly, $a_ {\nu \to i}$ has a gaussian
distribution with mean $0$ and a variance which is independent of the
indices $i$ and $\nu$ and that we denote by
$\overline{a^2}$. The rBP equations (\ref{rBP_a}) and (\ref{rBP_h})
relate these two variances:
\bea
\overline{h^2} &=& \alpha \overline{a^2}\\
 \overline{a^2} &=& \frac{q}{\left[1-\beta(1-q)\right]^2}
\eea

We thus obtain:
\be
q=\overline{\tanh^2 (\beta h)} = \int \frac{dh}{\sqrt{2\pi
    \Phi}} e^{-h^2/(2\Phi)}\ \tanh^2 (\beta h)\ ,
\label{replica1}
\ee
where :
\be
\Phi = \frac{\alpha q}{\left[1-\beta(1-q)\right]^2}  \ ,
\label{replica2}
\ee 
Eqs. (\ref{replica1},\ref{replica2}) are exactly the well known equations\cite{AGS85-2}
that allow to compute the spin glass order parameter $q$ in the spin
glass phase of the Hopfield model, in the replica-symmetric framework.

In the retrieval phase, the same reasoning can be applied starting
form the rBP-M equations (\ref{M_eq}-\ref{rBP_ret_h1}). One
finds:
\bea
q&=&\overline{\tanh^2 (\beta h+\xi M)} \\
M&=& \overline{\xi \tanh (\beta h+\xi M)}
\eea
where the overline denotes the average with respect to the field $h$,
which has a Gaussian distribution of variance $\Phi$, and the binary
variable $\xi$ which takes values $\pm 1$ with probability
$1/2$. These are precisely the equations obtained in the retrieval
phase with the replica method \cite{AGS85-2}.

\section{Algorithms : iterations and time indices}

Mean-field equations are usually solved by iteration, and interpreted
as message-passing algorithms. Turning a set of
mean-field equations into an iterative algorithm involves a certain
degree of arbitrariness concerning the way the equations are written
and the ``time indices'' concerning the update. A proper choice of
time indices may result in an algorithm with much better convergence
properties, as underlined for instance in \cite{bolthausen2014}, \cite{Zdeborova2015}.
Here we review the most natural choice for AMP iterations and their
consequences.

\subsection{rBP equations }
The rBP equations (\ref{rBP_a},\ref{rBP_h}) are usually iterated as
follows :
\bea
a_{\mu \to i}^{t+1} &=& \frac{1}{\sqrt N} \frac{
 \sum_{j(\neq i)} \xi_{j}^\mu \tanh(\beta h_{j \to \mu}^t)
}
{
1-(\beta/N)\sum_{j(\neq i)}  [1-\tanh^2(\beta h_{j\to\mu}^t)]
} \label{rBP_a_time}\\
h_{i\to \mu}^{t+2}&=& \sum_{\nu(\neq \mu)} \frac{\xi_i^\nu}{\sqrt{N}} a_{\nu
  \to i}^{t+1}
\label{rBP_h_time}
\eea

There exist various types of update schemes. One can distinguish two main
classes:
\begin{itemize}
\item
In the parallel update, starting from a configuration of the $h$
messages at time $t$, one computes all the $a$ messages using
(\ref{rBP_a_time}). Then one computes all the new $h$ messages at time
$t+2$ using
(\ref{rBP_h_time}), with the $a$ messages of time $t+1$ (therefore the
$h$-messages are defined at even times, the $a$-messages are defined at
odd times). In two
time-steps, all the
messages are updated. 
\item
In an update in series, one picks up a message at random (or, better,
one can use a random permutation of all messages to decide on the
sequence of updates), and one updates it using either
(\ref{rBP_a_time})-if the message is an $a$ message- or
(\ref{rBP_h_time}). In the case of random permutations, all messages
are updated after $2 N P$ time steps.
\end{itemize}
In the
parallel update scheme, one can easily follow the evolution in time of
the overlap $q^t$. Using (\ref{rBP_a_time},\ref{rBP_h_time}), one can
perform again the analysis of Sect.\ref{OP} keeping the time
indices. This gives:
\bea
q^{t+2} = \int \frac{dh}{\sqrt{2\pi
    \Phi^t}} e^{-h^2/(2\Phi^t)}\ \tanh^2 (\beta h)
\eea
where
\bea
\Phi^t= \frac{\alpha q^t}{\left[1-\beta(1-q^t)\right]^2}\ .
\eea
It is easy to see that these equations converge (to $q=0$) when $T>T_g=1+\sqrt{\alpha}$.

\subsection{TAP equations}
We can now repeat the previous derivation of the TAP 
equations keeping track of the time indices that were written in the previous
subsection. We keep here to the case of parallel update. 
Defining:
\bea
A_{\mu}^{t+1} &=&
\frac{1}{\sqrt N} \frac{
 \sum_{j} \xi_{j}^\mu \tanh(\beta h_{j \to \mu}^t)
}
{
1-(\beta/N)\sum_{j}  [1-\tanh^2(\beta h_{j\to\mu}^t)]
} 
\label{Adef_time}
\\
H_{i}^{t+2} &=& \sum_{\nu} \frac{\xi_i^\nu}{\sqrt{N}} a_{\nu
  \to i}^{t+1}
\label{Hdef_time}
\eea

One gets:
\bea
A_\mu^{t+1} &=&\frac{1}{1-\beta(1-q^{t})}\frac{1}{\sqrt N} \sum_{j}
\xi_{j}^\mu \tanh(\beta H_j^{t}
)-\frac{\beta(1-q^{t})}{1-\beta(1-q^{t})} A_\mu^{t-1}
\label{TAPdynA}
\\
H_i^{t+2}&=& \sum_{\nu} \frac{\xi_i^\nu}{\sqrt{N}}
A_\nu^{t+1}-\frac{\alpha}{1-\beta(1-q^{t})} \tanh(\beta H_i^{t})
\label{TAPdynH}
\eea
Equations (\ref{TAPdynA},\ref{TAPdynH}) give the algorithmic version
of TAP equations, used through a parallel iteration.
Again, the $A$ variables can be eliminated from these equations,
leaving the TAP equations written in terms of local field $H_i^t$ or
the magnetization $M_i^t=\tanh(\beta H_i^t)$. Defining $u^t=
\beta(1-q^{t})$, and $\tau=t/2$ we get:
\bea
H_i^{\tau+1}=\frac{1}{1-u^\tau}\left[
\sum_j J_{ij} M_j^{\tau}-\alpha M_i^{\tau} 
  -u^\tau H_i^\tau-\frac{\alpha u^\tau}{1-u^{\tau-1}}
  M_i^{\tau-1}\right]
\label{TAPdynMag}
\eea

This final form of the iterative algorithm corresponding to TAP
equation involves a kind of memory term (the polarization of neuron
$i$ at time
$\tau$ is obtained from the magnetizations at time $\tau-1$ and from
its magnetization at times $\tau-1$ and $\tau-2$), a phenomenon that
was first found in the context of TAP equations for the SK model
\cite{bolthausen2014}, and used in \cite{Zdeborova2015}.

This algorithm has many advantages. It
involves only $N$ fields, therefore its iteration is fast, and above
all it can develop a spontaneous polarization towards one of the
stored patterns (while in the rBP equations one would need to use a
different equation for each of the patterns).

\subsection{Numerical results}
The iteration of TAP equations (\ref{TAPdynA},\ref{TAPdynH}), or
equivalently their expression in terms of the $H$ fields only
(\ref{TAPdynMag}) is a fast algorithm for solving the Hopfield model
(in the sense of obtaining the local polarizations of the neuron-variables). We have tested it in
the retrieval phase, starting from a configuration with overlap $M_0$
with one randomly chosen pattern$ \mu_0$, which means that the initial spin
configuration $s_i^0$ satisfies
\be
M_0=\frac{1}{N}\sum_i s_i^0 \xi_i^{\mu_0}\ .
\ee

\begin{figure*}[!ht]
 \begin{center}
\hspace{-0.35cm}
 \includegraphics[width=0.45\textwidth]{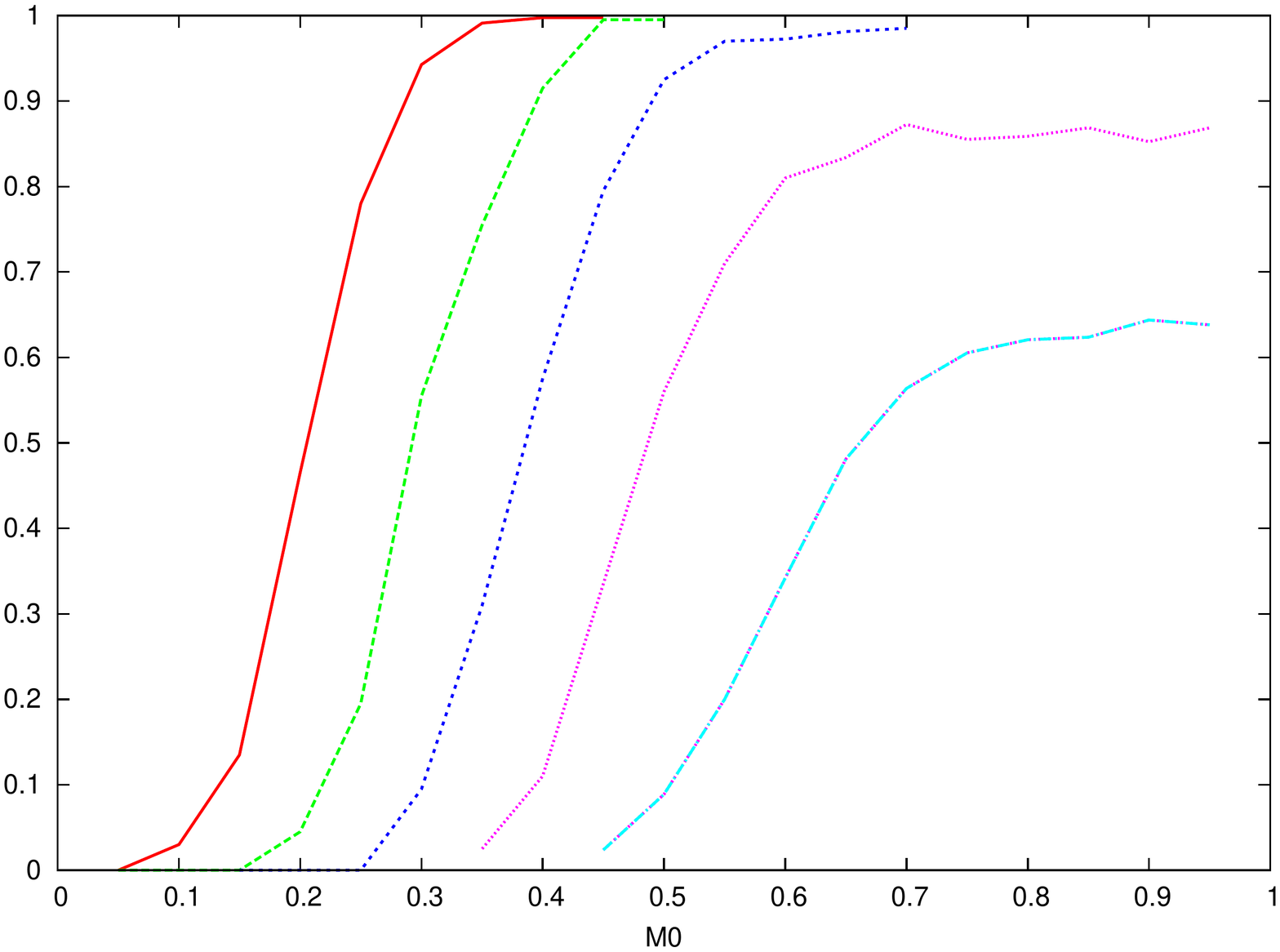}
\includegraphics[width=0.45\textwidth]{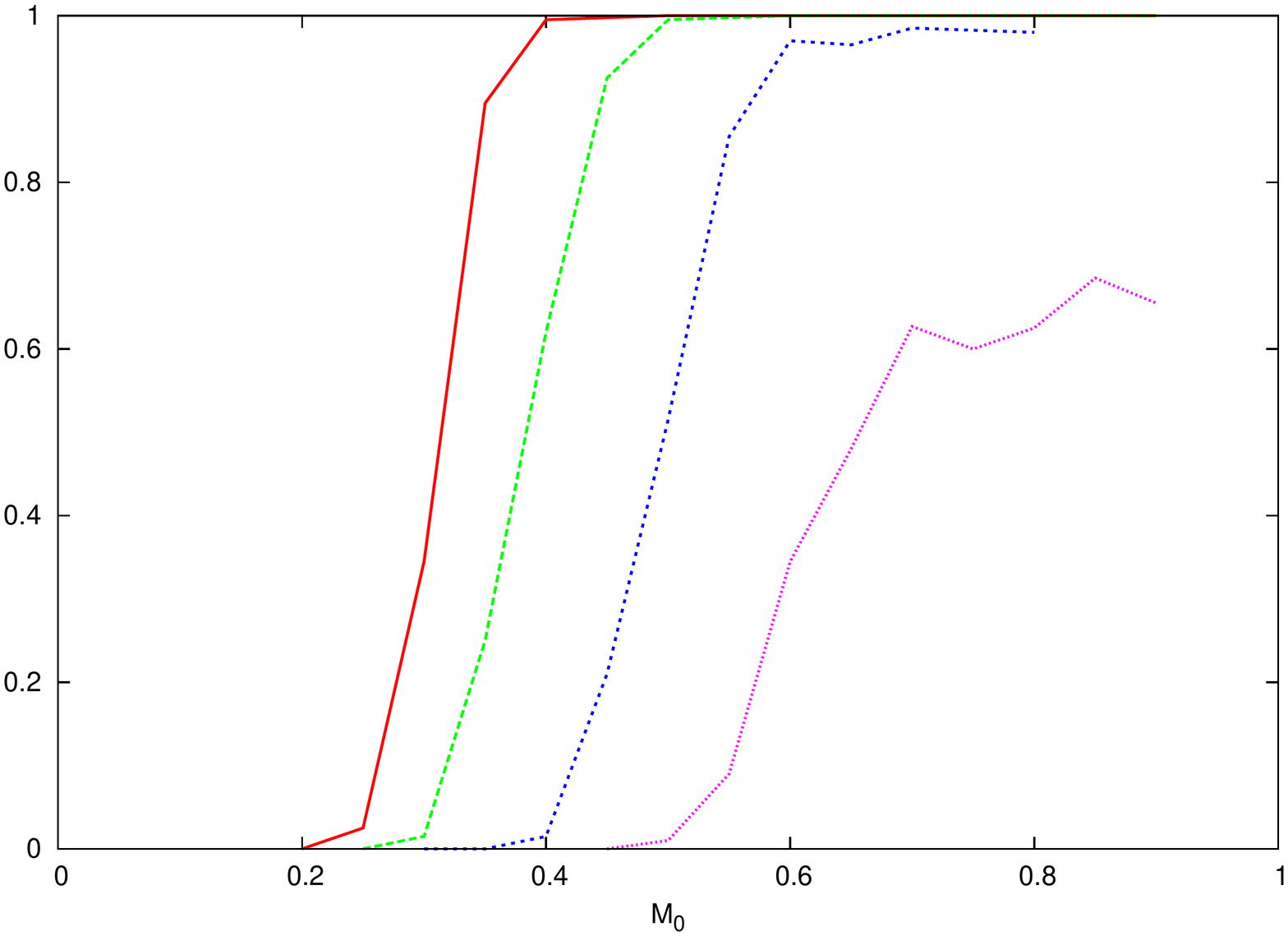}
\caption{Iteration of TAP equations using (\ref{TAPdynMag}). The
  probability of convergence is plotted versus the overlap to a
  randomly chosen initial pattern $\mu_0$. Left-hand figure: temperature
  $T=0.01$, simulation with $N=1000$ neurons and $P=40,60,80,100,120$
  patterns (from left to right). Right-hand figure:  temperature
  $T=0.3$, simulation with $N=1000$ neurons and $P=40,60,80,100$
  patterns (from left to right).}
\label{fig:iterTAP}
\end{center}
\end{figure*}

Fig.\ref{fig:iterTAP} shows the probability that the
iteration of these equations converges to a fixed point with a value
of overlap with $\mu_0$
larger than $.95$  (the
convergence is defined by
the fact that, in (\ref{TAPdynMag}), the average value of
$|H_i^{\tau+1}-H_i^\tau|<10^{-6}$). The simulations were carried out
with networks of $N=1000$ neurons. The maximal number of iterations
was fixed to $200$, but in practice we notice that when the algorithm
converges it does so in a few iterations, of order 10 to 20.

It should be noticed that the iteration of simpler versions of the mean-field
equations, either the naive mean-field equations or the SK-TAP
equations with the correct time indices of \cite{bolthausen2014}, also
converge when initialized in the same conditions. Actually the basin
of attraction for convergence to an overlap $>.95$ with the pattern is
larger for naive mean-field than it is for SK-TAP, and the one for
SK-TAP is larger than for the correct
Hopfield TAP equations. This is probably due to the fact, noticed in
\cite{KabashimaSaad2001}, that the fixed point reached by naive
mean-field is actually closer to the pattern than the fixed point
reached by SK-TAP, which is itself closer to the pattern than the one
obtained by iterating TAP equations. However, the TAP equations have
one major advantage : they give the values of the polarizations of
neurons which become exact in the thermodynamic limit.

\section{Modified Hopfield model: correlated patterns with 
  combinatorial structure}

From its definition (\ref{ene_Hopf}), the Hopfield model is a
 type of spin glass. It differs from the SK model by the
structure of couplings. In the SK model one draws each $J_{ij}$ (for $i<j$)
as an independent random variable with mean zero and variance
$1/N$. In the Hopfield model one builds the
$J_{ij}$ coupling constants as bilinear superposition of patterns, see
(\ref{Jdef}).  It turns out that this modification has a crucial
modification on the TAP equations. In the SK model, the TAP equations
are \cite{TAP}:
\bea
M_i =\tanh\left[ \beta \sum_{j (\neq i)} J_{ij} M_j -
\beta^2 (1-q) M_i \right]
\eea
The structure is the same in the Hopfield model, but the precise form
of the second term (the so-called Onsager reaction term) is
different. For an instructive comparison, it  is useful to rescale the interactions of the
Hopfield model in such a way that the variance of the couplings are
$1/N$, defining thus $ J_{ij}=\frac{1}{\sqrt{\alpha}}\frac{1}{N}\sum_\mu
  \xi_i^\mu \xi_j^\mu$. This simple rescaling can be absorbed in a
  rescaling of $\beta$, and our TAP equations (\ref{TAPmag1}) become,
  in this rescaled
  Hopfield model:
\bea
M_i = \tanh\left[ \beta \sum_{j(\neq i)} J_{ij} M_j -
\frac{\beta^2 (1-q)}{1-\beta(1-q)/\sqrt{\alpha}} M_i \right] 
\eea
Therefore the change of structure of the $J_{ij}$ random variables
leads to a modification of the TAP equations, where the Onsager term
acquires a denominator $1/[1-\beta(1-q)/\sqrt{\alpha}]$. Clearly, in
the large $\alpha$ limit one recovers the TAP equations of the SK
model, as it should be, since the correlations between the $J_{ij}$ become
irrelevant in this limit.

The fact that the TAP equations depend on the type of structure of the
couplings $J_{ij}$ poses a challenge for their use in practical
applications, where one does not really know the structure of these
couplings. One elegant way out consists in adapting the reaction term
to the concrete set of couplings to which one is applying the method
\cite{Tanaka1998,OpperWinther2001}. Our approach in the present paper
considers instead an alternative representation of the Hopfield model,
in which the visible neuron-variables interact with a hidden
layer of pattern-variables. In this expanded representation, the
couplings between the visible and the hidden units are nothing but the
patterns, which are independent random variables. Therefore the
message passing equations (BP, rBP, and eventually TAP) can be written
safely and give the result.

The standard results of the Hopfield model hold as long as
the $\xi_i^\mu$ are independent identically distributed (iid) random
variables with zero
mean and unit variance. We would like to test our approach by studying a generalization of the
Hopfield model in which the patterns are no longer independent random
variables. We shall study the case where the
patterns have a correlation, created from the following structure:
\be
\xi_i^\mu=\frac{1}{ \sqrt{\gamma N}} \sum_{r=1}^{\gamma N} u_i^r v_\mu^r
\label{xidef-combi}
\ee
where the $u_i^r$ are iid random variables drawn from a distribution $P_u$
with zero mean and unit variance, and the $v_\mu^r$ are iid random
variables drawn from a distribution $P_v$
with zero mean and unit variance.  Note that the scaling has been
chosen such that, in the large $\gamma$ limit, the pattern elements
$\xi_i^\mu$ become iid Gaussian random variables with unit variance, and one finds back the standard
Hopfield model.

We call the type of disorder generated by (\ref{xidef-combi}) a
combinatorial disorder. A natural case where it occurs is the
following: imagine that the patterns are built from a number $\gamma
N$ of possible features,
where the feature number $r$ is described by the neural activity
$u_i^r$. The variable $v_\mu^r$ encodes to what extent feature $r$
belongs to pattern $\mu$. For instance, using binary variables
$v_\mu^r=\pm 1$, one can interpret $v_\mu^r=1$ if and only if
feature $r$ belongs to feature $\mu$. Then the pattern $\mu$  on site $i$ is
( up to an overall constant), by the sum
of the features $r$ belonging to $\mu$. 

In combinatorial disorder, the random patterns  expressed as
(\ref{xidef-combi}) can be seen as a kind of  superposition of
features. This is in contrast with usual types of correlations that
were studied in previous years, like biased patterns, or
gaussian-distributed patterns with
a non-trivial correlation matrix. Obviously, the structure of combinatorial
disorder can be elaborated further and the features could become
themselves combination of subfeatures etc.

We shall now develop the mean-field equations for this modified model.

\subsection{Representation with hidden variables}

Using the representation (\ref{Z_def}), the partition function of the
modified  Hopfield
model with combinatorial disorder can be written as:
\be
Z= \sum_s \int \prod_\mu \frac{d \lambda_\mu e^{-\beta \lambda_\mu^2/2}}{\sqrt{2\pi\beta}}\ 
\exp\left[
\frac{\beta}{\sqrt{\gamma}}
\sum_{r=1}^{\gamma N}
 \left( \frac{\sum_{i}u_i^r s_i}{\sqrt{N}}  \right)
\left(\frac{\sum_{\mu}v_\mu^r \lambda_\mu }{\sqrt{N}} \right)
\right]\ .
\label{Z_def_mod}
\ee
It is useful to introduce the auxiliary variables 
\be
U^r = \frac{1}{\sqrt{N}}  \sum_{i}u_i^r s_i
\ee
and to use the representation
\be
1=\frac{\beta}{2 \pi i} \int dU^r d\hat U^r \exp\left[
\beta \hat U^r
    \left( \frac{1}{\sqrt{N}}
  \sum_{i} u_i^r s_i -U^r\right) 
\right]
\ee
where the auxiliary variable $\hat U^r$ is integrated in the complex
plane along the imaginary axis.

Similarly, we introduce the variable 
\be
V^r=\frac{1}{\sqrt{N}} \sum_{\mu} v_{\mu}^r \lambda_\mu
\ee
and write an integral representation in terms of an auxiliary variable
$\hat V^r$. 

This gives, up to some overall irrelevant constants:
\bea
Z&=& \sum_s \int \prod_\mu d \lambda_\mu\; \int\prod_r d \vec t^r  
e^{
-\frac{\beta}{2}\sum_\mu\lambda_\mu^2 + 
\beta \sum_{r=1}^{\gamma N}
\left(
+\frac{U^r V^r}{\sqrt{\gamma}}
-\hat U^r U^r- \hat V^r V^r
\right)
}
\nonumber \\
&&
\exp\left[
\frac{\beta}{\sqrt{N}}\sum_{r=1}^{\gamma N} \sum_{i=1}^N\hat
  U^r u_i^r s_i +
\frac{\beta}{\sqrt{N}}\sum_{r=1}^{\gamma N} \sum_{\mu=1}^{\alpha N}\hat
  V^r v_\mu^r \lambda_\mu
+\frac{\beta}{\sqrt{\gamma}}
     \sum_{r=1}^{\gamma N} U^r V^r
\right]
\label{Zaux}
\eea
where the variable $\vec t^r$ is: $\vec t^r=(\hat U^r,
U^r,\hat V^r,V^r)$, the integration element is   $d \vec t^r= d\hat U^r
dU^r d\hat V^r dV^r$, and the integrals overs $\hat U^r$ and $\hat
V^r$ run along the imaginary axis, while those over $U^r$ and $V^r$
are along the real axis. 

\begin{figure*}[!ht]
\begin{center}
\hspace{-0.35cm}
\includegraphics[width=0.95\textwidth]{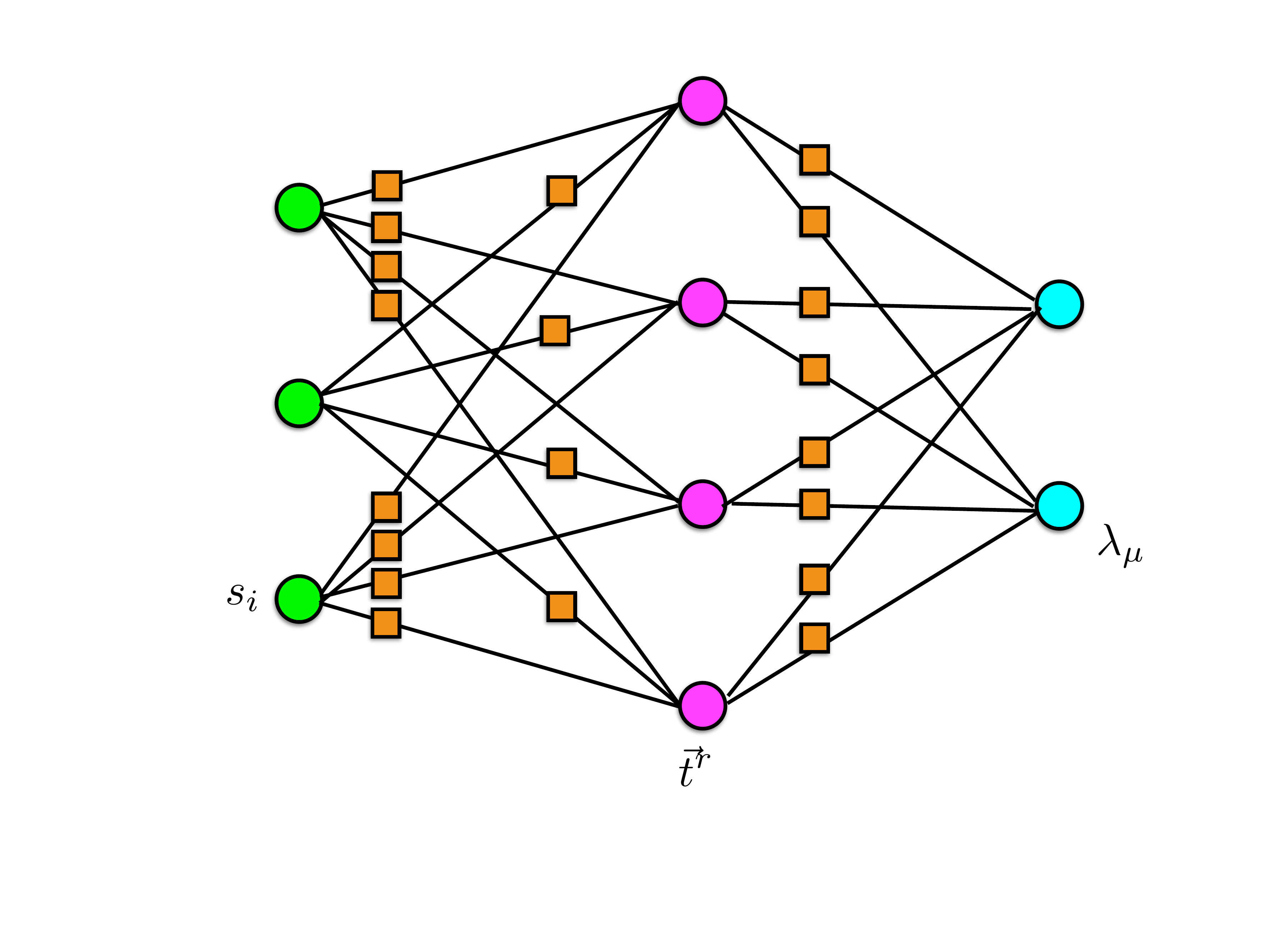}
\caption{Factor graph of the Hopfield model with combinatorial patterns, in the representation
  using visible spin-variables ($s_i$, left-layer green circles),
 hidden  pattern-variables ($\lambda_\mu$, right-layer blue circles) and
 hidden feature-variables (middle-layer purple circles). There exist interaction
  factors (squares) between each pair of variables belonging to two
  consecutive layers.}
\label{fig:factorgraph2}
\end{center}
\end{figure*}

The representation (\ref{Zaux}) contains three types of variables:
\begin{itemize}
\item The $N $ visible  neuron-variables $s_i$.
\item The $\alpha N$ pattern-variables $\lambda_\mu$, which are hidden
  variables.
\item The $\gamma N$ ``feature-variables'' $\vec t^r$, which build a new
  layer of hidden variables, interacting with the other two layers.
\end{itemize}
Figure \ref{fig:factorgraph2} shows the factor graph for this
problem. 

\subsection{Belief propagation}

Writing the BP equations for the model (\ref{Zaux}) is a standard (but
lengthy) exercise that goes along exactly the same lines as before. It
involves 8 types of messages running along the edges of the
factor graph shown in Fig. \ref{fig:factorgraph2}. These messages are:
\bea
&m&_{i\to r} (s_i) \ ,\ \hat m_{r \to i} (s_i)\ ,\ m_{\mu \to
  r}(\lambda_\mu) \ ,\ \hat m_{r\to \mu}(\lambda_\mu)\\
&\hat m&_{i\to r}(\vec t^r) \ ,\ m_{r\to i}(\vec t^r) \ ,\  m_{r
  \to \mu}(\vec t^r) \ ,\ \hat m_{\mu \to r}(\vec t^r) 
\eea
Here and in the following, the letters $i,j$ are indices of the
neuron-variables running from $1$ to $N$, the letters $r,s$ are indices of the
feature-variables running from $1$ to $\gamma N$, and the letters $\mu,\nu$ are indices of the
pattern-variables running from $1$ to $\alpha N$. Each message is a function of the argument which is
written in parenthesis.

We shall not write explicitly the BP equations, but proceed directly
to the rBP ones, which can be expressed in terms of the messages $h_{i\to
  r},a_{\mu \to r},c_{\mu \to r}$ defined from:
\bea
&&m_{i\to r}(s_i)\normeq e^{h_{i\to r} s_i}\\
&&\int d\lambda_\mu m_{\mu\to r}(\lambda_\mu)\; \lambda_\mu= a_{\mu\to
  r}\\
&&\int d\lambda_\mu m_{\mu\to r}(\lambda_\mu)\; \lambda_\mu^2= c_{\mu\to
  r}+a_{\mu\to
  r}^2
\eea

They are related through the following set of equations:
\bea
h_{i \to r}&=&\frac{1}{\sqrt{N}}\sum_{s(\neq r)} u_i^s f
\left(p_{s
    \to i}, d_{s \to i},\pi_s,\delta_s\right)
\label{hcavmod-def}\\
c_{\nu \to r}&=&\frac{1}{\beta}\left[1-\frac{1}{N}\sum_{s(\neq
    r)}(v_\nu^s)^2 \phi' (p_s,d_s,\pi_{s\to \nu}, \delta_{s\to
    \nu})\right]^{-1}\label{ccavmod-def}\\
a_{\nu \to r}&=&\beta c_{\nu \to r}\; \frac{1}{N}\sum_{s(\neq
    r)} v_\nu^s \phi(p_s,d_s,\pi_{s\to \nu}, \delta_{s\to
    \nu})\label{acavmod-def}
\eea
where:
\bea
p_r &=& \frac{1}{\sqrt{N}} \sum_i u_i^r \tanh(\beta h_{i\to r})\\
d_r &=& \frac{1}{{N}} \sum_i (u_i^r)^2 (1- \tanh^2(\beta h_{i\to
  r}))\\
\pi_r&=&\frac{1}{\sqrt{N}} \sum_\nu v_\nu^r a_{\nu \to r}\\
\delta_r&=&\frac{1}{{N}} \sum_\nu (v_\nu^r)^2 C_{\nu \to r}
\eea 
and 
\bea
p_{r \to j}&=& \frac{1}{\sqrt{N}} \sum_{i(\neq j)} u_i^r \tanh(\beta h_{i\to r})\\
d_{r \to j}&=& \frac{1}{{N}} \sum_{i(\neq j)} (u_i^r)^2 (1- \tanh^2(\beta h_{i\to
  r}))\\
\pi_{r \to \mu}&=&\frac{1}{\sqrt{N}} \sum_{\nu (\neq \mu)} v_\nu^r a_{\nu \to r}\\
\delta_{r \to \mu}&=&\frac{1}{{N}} \sum_{\nu (\neq \mu)} (v_\nu^r)^2 c_{\nu \to r}
\eea 
The functions $f,\phi,\phi'$ are functions of four variables defined
as:
\bea
f(p,d,\pi,\delta) &=&\langle \hat U \rangle\\
\phi(p,d,\pi,\delta) &=&\langle \hat U^2 \rangle-\langle \hat U
\rangle^2 \\
\phi'(p,d,\pi,\delta) &=&\frac{\partial}{\partial
  \pi}\phi(p,d,\pi,\delta)
\eea
where the expectations are taken with the following measure over $\vec
t=(U,V,\hat U,\hat V)$:
\be
\exp\left[
\beta\left(
-\hat U U -\hat V
    V+\frac{1}{\sqrt \gamma} U V+p \hat U + \pi \hat
    V
\right)
+\frac{\beta^2}{2}\left(
d \hat U^2+ \delta \hat V^2
\right)
\right]
\ee
An explicit computation shows that
\bea
f(p,d,\pi,\delta)&=& \frac{1}{1-d\delta \beta^2/\gamma}
\left(
\frac{\beta\delta}{\gamma} p + \frac{1}{\sqrt\gamma}\pi
\right)
\\
\phi(p,d,\pi,\delta)&=& \frac{1}{1-d\delta \beta^2/\gamma}
\left(
\frac{1}{\sqrt\gamma} p + \frac{\beta d}{\gamma}\pi
\right)
\\
\phi'(p,d,\pi,\delta)&=&\frac{1}{1-d\delta \beta^2/\gamma}
\left(\frac{\beta d}{\gamma}\right)
\eea

\subsection{TAP equations}
It turns out that the TAP equations can be written in terms of local
quantities associated with each of the variable in the factor graph: starting from (\ref{hcavmod-def},\ref{ccavmod-def},\ref{acavmod-def}) we define
\bea
H_i&=& \frac{1}{\sqrt{N}}\sum_s u_i^s f(p_{s \to i},d_{s\to
  i},\pi_s,\delta_s)\\
C_{\nu }&=&\frac{1}{\beta}\left[1-\frac{1}{N}\sum_{s}(v_\nu^s)^2 \phi'
  (p_s,d_s,\pi_{s\to \nu}, \delta_{s\to
    \nu})\right]^{-1}\\
A_{\nu }&=&\beta c_{\nu \to r}\; \frac{1}{\sqrt N}\sum_{s} v_\nu^s \phi(p_s,d_s,\pi_{s\to \nu}, \delta_{s\to
    \nu})
\eea
We first notice that, in the thermodynamic limit, $C_\nu=C$ becomes
independent of $\nu$, and we can also safely approximate $c_{\nu\to
  r}=C$, the correcting terms being irrelevant. Similarly, we notice
that $d_r$ becomes $r$-independent,
\be
d_r=1-q=1-\frac{1}{N}\sum_i \tanh^2(\beta H_i) \ ,
\ee
and $\delta_r$ becomes $r$-independent:
\be
\delta_r=\alpha C\ .
\ee
The equation for $C$ can be obtained from (\ref{ccavmod-def}):
\be
\frac{1}{C}=\beta-\frac{\beta^2(1-q)}{1-C\alpha\beta^2
  (1-q)/\gamma}
\ee
and gives
\bea
C=\frac{\gamma}{2 \alpha \beta^2
  (1-q)}&&\left[1-\beta(1-q)(1-\alpha/\gamma)\right.\nonumber\\
&&
-\left. \sqrt{(1-\beta(1-q)(1+\alpha/\gamma))^2-4\alpha\beta^2(1-q)^2/\gamma}\right]
\label{Cdef}
\eea
$C$ is nothing but the variance of the local-fields $A_\nu$ for each
pattern-variable. When $\gamma\to \infty$ one finds back that
$C=(1/\beta)1/[1-\beta(1-q)]$, which is the expression found in the
Hopfield model, as it should. 

Defining:
\bea
\hat f(p,\pi)&=& f(p,1-q,\pi,\alpha C)= \frac{1}{1-C\frac{\alpha}{\gamma}\beta^2
  (1-q)}\left[
\frac{\alpha \beta C}{\gamma} p+\frac{1}{\sqrt\gamma}\pi
\right]\label{hatfdef}\\
\hat \phi(p,\pi)&=& \phi(p,1-q,\pi,\alpha C)= \frac{1}{1-C\frac{\alpha}{\gamma}\beta^2
  (1-q)}\left[\frac{1}{\sqrt\gamma} p+
  \frac{\beta(1-q)}{\gamma}\pi\right]
\label{hatphidef}
\eea
We can write the TAP equations:
\bea
H_i&=& \frac{1}{\sqrt{N}}\sum_s u_i^s \; \hat f(p_s,\pi_s)-
\frac{\alpha\beta C}{1-C\frac{\alpha}{\gamma}\beta^2 (1-q)}
\tanh(\beta H_i)
\label{TAPmod1}\\
A_{\nu }&=& \frac{1}{N}\sum_{s} v_\nu^s \; \hat \phi(p_s,\pi_{s})\\
p_r &=& \frac{1}{\sqrt{N}} \sum_i u_i^r \; \tanh(\beta
H_i)-\frac{\beta(1-q)}{\sqrt\gamma}\frac{1}{\sqrt{N}} \sum_\nu v_\nu^r\;
A_\nu\\
\pi_r&=&\frac{1}{\sqrt{N}} \sum_\nu v_\nu^r\; 
A_\nu-\frac{\alpha\beta C}{\sqrt{\gamma}}\frac{1}{\sqrt{N}} \sum_i
u_i^r \; \tanh(\beta
H_i)\label{TAPmod4}
\eea
Equations (\ref{TAPmod1}-\ref{TAPmod4}), together with the definitions
(\ref{Cdef},\ref{hatfdef},\ref{hatphidef}) give the closed set of TAP
equations relating the $N(1+\alpha+2\gamma)$  local fields
$H_i,A_\nu,p_r,\pi_r$.

It is interesting to notice that, due to the linear structure of these
equations, the variables $p_r,\pi_r$ can be eliminated explicitly,
leading to a set of equations that relate only the fields on the
site-variables, $H_i$, and those on the pattern-variables, $A_\mu$:
\bea
H_i&=& \sum_{\nu} \frac{\xi_i^\nu}{\sqrt{N}}
A_\nu-\frac{\alpha\beta C}{1-C\frac{\alpha}{\gamma}\beta^2 (1-q)}
\tanh(\beta H_i)\\
A_\mu&=&\frac{1}{\sqrt N} \sum_{j} \xi_{j}^\mu \tanh(\beta H_j )
\eea
These TAP equations are similar to the ones of the Hopfield model,
with a modified form of the Onsager reaction term. Again, because of
their linear structure in $A_\mu$, these variables can be eliminated,
giving a set of TAP equation connecting only the local fields on the
visible neuron-variables:
\bea
H_i=\frac{1}{N} \sum_j J_{ij} \tanh(\beta H_j) -
\frac{\alpha\beta C}{1-C\frac{\alpha}{\gamma}\beta^2 (1-q)}\tanh(\beta H_i)
\eea
Again, the only modification due to the combinatorially correlated
patterns is the value of the Onsager reaction term.
Notice that, in the large $\gamma$ limit, we get back the usual TAP
equation of the Hopfield model.

We have derived four versions of the mean-field equations for this
modified Hopfield: the rBP equations which relate messages that are propagated on the edges of the
factor graph, and three versions of the TAP equations, one set of
``expanded'' equations which
relate local quantities associated with each variable node of the factor graph, a
second one, ``intermediate'',  which relates the local fields of the neuron-variables and
the pattern-variables, and finally this last one that relates only the
local fields on the neuron-variables. Which one is more useful remains
to be investigated. The rBP equations should be studied statistically,
and give the solution for the thermodynamic properties of this
modified Hopfield model using the cavity method. The schedule of
update of TAP equations is probably crucial, and working out the
correct time indices for algorithmic purpose should go through the
expanded version of the equations. It should also be kept in mind
that, in general RBMs, the hidden variables are in general not
Gaussian-distributed, and in such cases the simplification of TAP
equations does not occur (see the next section). Therefore in general
the correct form of TAP equations can be obtained only in their
expanded form. This shows the importance of using multilayered networks.

\section{A few remarks on more general restricted Boltzmann machines}
It is easy to generalize the Hopfield model in order to describe a
general  RBM. We shall give here the general form of BP, rBP and TAP
equations. Similar results have been obtained recently in
information-theoretic approaches to matrix factorization
\cite{Deshpande2014,Lesieur2015}, but they generally address a form
of ``planted'' problem where specific simplifications take place
\cite{KrzakalaZdeborova09,Zdeborova2015}. We give here the general form of the equations.

Using the same notations as before, we consider a system
of $N$ spin-variables $s_i$ and $P$ pattern-variables $\lambda_\mu$,
described by a probability distribution:
\bea
P(\{s_i\},\{\lambda_\mu\})&=&\frac{1}{Z} \prod_i \tilde \rho(s_i) \prod_\mu
\rho(\lambda_\mu)\nonumber \\ &&
\exp\left[\beta\left(\sum_i \tilde h_i s_i+\sum_\mu h_\mu \lambda_\mu+
  \sum_{\mu,i}\frac{\xi_i^\mu}{\sqrt{N}} s_i \lambda_\mu\right)
\right]\ .
\eea
With respect to the usual Hopfield model, three modifications have
been introduced:
\begin{itemize}
\item
The local measure on the spin variables is $\tilde \rho(s)$. In the
Hopfield model one considers
$\tilde \rho(s)=(1/2)(\delta_{s,1}+\delta_{s,-1})$, but more general
distributions can be studied as well.
\item
The local measure on the pattern variables is $ \rho(\lambda)$. In the
Hopfield model one considers
$\rho(\lambda)=(1/\sqrt{2\pi \beta}) \exp(-\beta \lambda^2/2)$, but more general
distributions can be studied as well.
\item
We introduce local fields $\tilde h_i$ and $h_\mu$, which will make it
possible to compute correlation functions through linear response,
using for instance $\langle s_i s_j\rangle=\partial \langle
s_i\rangle/\partial \tilde h_j$.
\end{itemize}

The  BP equations are:
\bea
m_{i\to\mu}(s_i) &\normeq& \tilde \rho(s_i)\; e^{\beta \tilde h_i s_i} \; \prod_{\nu(\neq \mu)} \hat
m_{\nu \to i} (s_i)  \label{genBP1}\\
\hat m_{\mu \to i}  (s_i) &\normeq&  \int d\lambda_\mu 
m_{\mu\to i}(\lambda_\mu) \exp\left(  (\beta/\sqrt{N}) \xi_i^\mu s_i
  \lambda_\mu\right)\label{genBP2}\\
\hat m_{i\to\mu}(\lambda_\mu)&=& \int d s_i\; m_{i\to \mu}(s_i)
e^{\beta\xi_i^\mu/\sqrt{N}}
\label{genBP3}\\
m_{\mu\to i}(\lambda_\mu) &\normeq&\rho(\lambda_\mu)\; e^{\beta
  h_\mu \lambda_\mu}\; \prod_{j(\neq
  i)} \hat m_{j\to \mu}(\lambda_\mu)\label{genBP4}
\eea

In order  to write the rBP equations, we need to understand the
scaling of the variables. In particular, we have seen in the
Hopfield model that, in the retrieval phase, one of the variables
$\lambda_\mu$ may become very large (of order $\sqrt{N}$), signalling a
polarization towards this pattern. The possibility of such a phenomenon clearly depends on
the measures $\tilde \rho(s)$ and $\rho(\lambda)$. In the Hopfield
case, $\rho(\lambda)$ is a Gaussian. This means that the response of
the pattern-variable $\lambda$ to a local field $h$ is a linear
function of $h$. This allows the variable $\lambda_\mu$ to grow to
very large values. In contrast, in many applications of RBMs, one uses
 variables with a bounded range of values. For instance, if
$\rho(\lambda_\mu)$ vanishes outside an interval $[-C,C]$, then the
response of the variable $\lambda_\mu$ is a non-linear, sigmoid-shaped
function of the local field, and the
condensation cannot occur.

One opposite case would be the one when both  $\tilde \rho(s)$ and
$\rho(\lambda)$ are Gaussian. It is then clear that, at low
temperatures, the spins will acquire spontaneous polarization in the
direction of the eigenvector of the $J$ matrix (Eq.\ref{Jdef}) with
largest eigenvalue. Both the spin-variables and the pattern-variables
condense in this case.

We shall write here the rBP equations assuming that there is no
condensation. As for the coupling variables $\xi_i^\mu$, we suppose that
they are iid variables with zero mean and a finite variance.

Following standard procedures like those used in
\cite{BayatiMontanari10,Rangan10,Rangan10b,KrzakalaPRX2012,KrzakalaMezard12},
the messages $m_{\mu \to i}(\lambda_\mu)$ and $ m_{i\to\mu}(s_i)$ are
parameterized in terms of their first two moments. Generalizing
(\ref{adef},\ref{cdef}), we define:
\bea
a_{\mu\to i} &= & \int d\lambda_\mu 
m_{\mu\to i}(\lambda_\mu) \lambda_\mu \label{adef2}\\
c_{\mu\to i} &= & \int d\lambda_\mu 
m_{\mu\to i}(\lambda_\mu) \lambda_\mu^2- a_{\mu\to i}^2\label{cdef2}\\
\tilde a_{i\to \mu} &=& \int d s_i m_{i\to\mu}(s_i) s_i\\
\tilde c_{i\to \mu} &=& \int d s_i m_{i\to\mu}(s_i) s_i^2- \tilde a_{i\to \mu}^2
\eea

The rBP equations relating these four types of messages can be written in terms of the following four
functions of two real variables.

Considering a spin-variable $s$ with local measure 
\be
\tilde P(s)=\frac{1}{\tilde z}\tilde \rho(s) e^{u s +(v/2)s^2}
\ee
we define 
\bea
\tilde f(u,v)&=& \int ds \tilde P(s) s\\
\tilde f'(u,v) &=& \frac{\partial}{\partial u} \tilde f(u,v)=  \int ds \tilde P(s) s^2- \tilde f_a(u,v)^2
\eea

Considering a pattern-variable $\lambda$ with local measure 
\be
 P(\lambda)=\frac{1}{ z} \rho(\lambda) e^{u \lambda +(v/2) \lambda^2}
\ee
we define 
\bea
f(u,v)&=& \int d \lambda  P(\lambda) \lambda\\
f'(u,v) &=& \frac{\partial}{\partial u} f(u,v)= \int d \lambda  P(\lambda) \lambda^2-  f_a(u,v)^2
\eea

The rBP equations can then be written as
\bea
a_{\mu\to i}^{t+1}&=&f\left(\beta h_\mu+
\frac{\beta}{\sqrt N} \sum_{j(\neq i)} \xi_j^\mu \tilde a_{j\to \mu}^t\;
, \;
\frac{\beta^2}{ N} \sum_{j(\neq i)} (\xi_j^\mu)^2 \tilde c_{j\to \mu}^t
\right)\\
c_{\mu\to i}^{t+1}&=&f'\left(\beta h_\mu+
\frac{\beta}{\sqrt N} \sum_{j(\neq i)} \xi_j^\mu \tilde a_{j\to \mu}^t\;
,\; 
\frac{\beta^2}{ N} \sum_{j(\neq i)} (\xi_j^\mu)^2 \tilde c_{j\to \mu}^t
\right)\\
\tilde a_{i \to\mu}^{t+2}&=&\tilde f\left(\beta \tilde h_i+
\frac{\beta}{\sqrt N} \sum_{\nu(\neq \mu)} \xi_i^\nu  a_{\nu\to i}^{t+1}\; , \;
\frac{\beta^2}{ N} \sum_{\nu(\neq \mu)} (\xi_i^\nu)^2  c_{\nu\to i}^{t+1}
\right)\\
\tilde c_{i \to\mu}^{t+2}&=&\tilde f'\left(\beta \tilde h_i+
\frac{\beta}{\sqrt N} \sum_{\nu(\neq \mu)} \xi_i^\nu  a_{\nu\to i}^{t+1} \;
, \;
\frac{\beta^2}{ N} \sum_{\nu(\neq \mu)} (\xi_i^\nu)^2  c_{\nu\to i}^{t+1}
\right)
\eea
where we have reintroduced the time indices corresponding to a
parallel update of these equations.

One gets the TAP equations using the same method as before.
 In the large $N$
limit the messages depend only weakly on the index of arrival. Writing 
\bea
a_{\mu\to i}\simeq A_\mu \ \ ; \ \ c_{\mu\to i}\simeq C_\mu \ \ ; \ \ \tilde
a_{i \to\mu}\simeq \tilde A_i  \ \ ;\ \   \tilde c_{i \to\mu}\simeq C_i
\eea
and expanding the leading correction terms, one obtains:
\bea
A_\mu^{t+1}&=&f(U_\mu^t,V_\mu^t)
\\
C_\mu^{t+1}&=&f'(U_\mu^t ,V_\mu^t)  \\
\tilde A_i^{t+2}&=&\tilde f(\tilde U_i^{t+1},\tilde V_i^{t+1})
\\
\tilde C_i^{t+2} &=& \tilde f'(\tilde U_i^{t+1}, \tilde V_i^{t+1})
\eea
where :
\bea
U_\mu^t&=&\beta h_\mu+\frac{\beta}{\sqrt N}\sum_i \xi_i^\mu \tilde A_i^t
-A_\mu^{t-1} \frac{\beta^2}{N}\sum_i (\xi_i^\mu)^2\;
  \frac{\partial \tilde f}{\partial \tilde U}(\tilde U_i^{t-1}, \tilde
  V_i^{t-1})
\\
V_\mu^t&=& \frac{\beta^2}{N}\sum_i (\xi_i^\mu)^2 \; \tilde C_i^t\\
\tilde U_i^{t+1}&=&\beta \tilde h_i+\frac{\beta}{\sqrt N}\sum_\mu\xi_i^\mu
A_\mu^{t+1}
-\tilde A_i^t \frac{\beta^2}{N}\sum_\mu (\xi_i^\mu)^2\;
  \frac{\partial  f}{\partial  U}( U_\mu^t , 
  V_\mu^t)
\\
\tilde V_i^{t+1}&=& \frac{\beta^2}{ N}\sum_\mu (\xi_i^\mu)^2\;
C_\mu^{t+1}
\eea
Notice that, in general, when $\tilde \rho(s)$ and $\rho(\lambda)$ are
non-gaussian, the functions $f$ and $\tilde f$ are nonlinear functions
of $u$, and therefore one cannot easily eliminate one of the
variables, as was done in the Hopfield model. This means that the
correct form of  TAP
equations require working on the bipartite graph with the two layers
of variables, visible and hidden. 

\section{Concluding remarks}
We have seen that the correct mean-field ''TAP''  equations in the Hopfield model
can be written most easily by introducing a layer of hidden variables,
the pattern-variables, which interact with the neuron-variables. In
the Hopfield model, the local fields associated with the hidden
variables can be eliminated and one remains with TAP equations that
are similar to those of general spin glasses, differing only in the
detailed form of the Onsager reaction term. However, when
one deals with RBMs which generalize the Hopfield model with 
non-Gaussian hidden variables, the representation with the hidden
layer is necessary. 

In the case where the patterns to be memorized have correlations based
on a combinatorial structure, the TAP equations involve one extra
layer of hidden variables, and with a deeper structure of
correlations, extra hidden layers would be added. 

 We believe that combinatorial disorder is
actually an essential ingredient that is likely to be present in real
data. In this respect, it is striking that the correct treatment of
mean-field theory in RBMs with combinatorial disorder leads naturally
to the appearance of layers of hidden variables. The present study of the Hopfield model is a kind
of first test of this idea, which we hope could lead to a better
understanding of the role of multilayered structures in practical
applications of neural networks.

The present work calls for some further developments in several directions:
\begin{itemize}
\item It will be interesting to study how the TAP estimates for the
magnetizations (and those for the correlation functions that are
inferred through linear response) can be turned into efficient
algorithms for unsupervised learning, along the lines of
\cite{KappenRodriguez1998,Tanaka1998,HuangToyoizumi2015,GabrieTramelKrzakala2015,TramelManoel_etal2016}.
In this respect, it is interesting to be able to study controlled
problems. We think that the Hopfield model with
combinatorial-correlated patterns can be used as an interesting
teacher to generate data, i.e. patterns of neural activity, that can
be used in the training of a ``student'' Hopfield network.
\item The modified Hopfield model with combinatorial-correlated patterns
is interesting in itself. It would be interesting to study its
thermodynamics both with replicas and with the cavity method, through
a statistical analysis of the rBP equations.
\end{itemize}

\section*{Acknowledgments} 
It is a pleasure to thank C. Baldassi, F. Krzakala, L. Zdeborov\' a
and R. Zecchina for
interesting discussions related to the subject of this note.

\bibliography{refs}
\end{document}